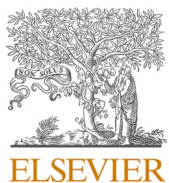
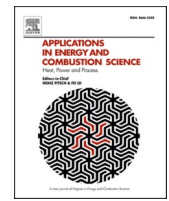

# Application of dense neural networks for manifold-based modeling of flame-wall interactions

Julian Bissantz [a], Jeremy Karpowski [a,1], Matthias Steinhausen [a], Yujuan Luo [a], Federica Ferraro [a,*], Arne Scholtissek [a], Christian Hasse [a], Luc Vervisch [b]

[a] *Technical University of Darmstadt, Department of Mechanical Engineering, Simulation of reactive Thermo-Fluid Systems, Otto-Berndt-Str. 2, 64287 Darmstadt, Germany*
[b] *CORIA - CNRS, Normandie Université, INSA de Rouen, Technopole du Madrillet, BP 8, Saint-Etienne-du-Rouvray 76801, France*

ABSTRACT

Artifical neural networks (ANNs) are universal approximators capable of learning any correlation between arbitrary input data with corresponding outputs, which can also be exploited to represent a low-dimensional chemistry manifold in the field of combustion. In this work, a procedure is developed to simulate a premixed methane-air flame undergoing side-wall quenching utilizing an ANN chemistry manifold. In the investigated case, the flame characteristics are governed by two canonical problems: the adiabatic flame propagation in the core flow and the non-adiabatic flame- wall interaction governed by enthalpy losses to the wall. Similar to the tabulation of a Quenching Flamelet-Generated Manifold (QFM), the neural network is trained on a 1D head-on quenching flame database to learn the intrinsic chemistry manifold. The control parameters (i.e. the inputs) of the ANN are identified from thermo-chemical state variables by a sparse principal component analysis (PCA) without using prior knowledge about the flame physics. These input quantities are then transported in the coupled CFD solver and used for manifold access during simulation runtime. The chemical source terms are corrected at the manifold boundaries to ensure bounded- ness of the thermo-chemical state at all times. Finally, the ANN model is assessed by comparison to simulation results of the 2D side-wall quenching (SWQ) configuration with detailed chemistry and with a flamelet-based manifold (QFM).

## 1. Introduction

In the transition towards sustainable combustion technologies, numerical simulations play a crucial role for the rapid design of carbon-neutral or carbon-free combustion systems for power generation and transportation. While highly-resolved detailed chemistry (DC) simulations are essential to understand physical phenomena and can serve as the basis for model development and validation, their application to practical combustors is often unfeasible due to prohibitive computational costs. Consequently, there exists a demand for accurate and numerically efficient chemistry reduction approaches. One option are reduced-order models utilizing chemistry manifolds [1–4]. These methods are based on the tabulation of pre-calculated thermo-chemical states which are pa- rameterized and accessed by control variables. The control variables are then usually transported in the CFD simulation and used to retrieve the thermo-chemical state from the manifold during simulation runtime. Given a suitable manifold is used, these methods combine the high accuracy of a detailed chemistry (DC) simulation with low computational costs. However, for complex configurations involving a significant number of different combustion phenomena (e.g., multi-phase combustion, pollutant formation modeling or flame-wall interactions), more control variables, thus, additional manifold dimensions are required. The memory demand to store the man- ifold quickly becomes intractable considering memory-per-core limitations on high-performance computers. Additionally, the tabulation of the manifold becomes increasingly difficult since the control variables are often not linearly independent, which complicates a non-overlapping data arrangement and efficient manifold access.

An emerging alternative is data-driven modeling, or machine learning (ML), using artificial neural networks (ANN). Neural networks are universal approximators capable of learning any correlation between arbitrary input data (control variables) with corresponding outputs. This holds true even if the control variables are not linearly independent. This property of ANNs can be exploited to represent a chemistry manifold. An ANN is memory efficient and the storage size changes only slightly with the number of control variables. Furthermore, ANNs profit from an enormous performance increase on specialized accelerator hardware, which is one focus of current computational hardware development.

The first ANN application for chemistry modeling in a Large Eddy Simulation (LES) was performed by Flemming et al. [5], who used the approach to model the Sandia Flame D. The authors achieved a memory reduction by three orders of magnitude with a minimal increase in computational time in comparison with a tabulated manifold approach. Ihme et al. [6] focused on the optimization of ANN architectures for each output variable using a generalized pattern search. These models were subsequently applied in an LES of the Sydney bluff-body swirl-stabilized SMH1 flame. Again, the ANN showed a comparable accuracy to a tabulated manifold with acceptable computational overhead for LES applications. Recently, several investigations have been carried out applying machine learning in reactive flow simulations for manifold representation [7–13], turbulence-chemistry interactions [14–17] and modeling chemical kinetics [18–22]. This was made possible by the development of open-source deep learning frameworks that followed the breakthrough of deep learning in the field of computer vision in the past decade. Although the training of ANNs has been simplified by these frameworks, many works have reported a lack of model accuracy in case input values approach the manifold boundaries. This applies in particular to non-linear quantities, such as chemical source terms. In a direct numerical simulation (DNS) of a turbulent syngas oxy-flame, Wan et al. [23,24] trained additional neural networks for a normalized oxygen mass fraction $Y_{O_2} > 0.9$ to improve the model accuracy close to the unburnt conditions, where low, but non-zero, reaction rates occured. Similarly, Ding et al. [10] used a multiple multilayer perceptron (MMP) approach, where additional models were trained on increasingly smaller intervals around zero, where the relative error of the prediction is large compared to the absolute error. During the simulation, a model cascade is employed, where the decision which output is used is based on the output value of the previous model. This methodology was applied in LES of the Sandia flame series. Another approach to increase the overall prediction accuracy of ML models is to divide the manifold using clustering algorithms such as self-organizing maps [8,12,25,26] or k-means clustering [27] and to use different ANNs for the prediction of the subsets of the manifold. However, in some cases, hundreds of networks had to be trained [8,25], which introduces additional overhead for the model selection during simulation runtime. For thorough overviews of machine learning approaches in the context of combustion, the reader is referred to Zhou et al. [28] and Ihme et al. [29].

While there exist many coupled simulations with models based on tabulated reduced-order manifolds, the literature on ML-based manifold modeling is still scarce. First results have been encouraging, but several challenges, such as the handling of non-linear terms, a robust feature selection (i.e. suitable control variables), or the required number of networks for an accurate manifold representation, are active areas of research.

In this work, a ML model based on dense neural networks (DNN) is coupled to a CFD solver and utilized for the 2D simulation of a laminar premixed methane-air flame undergoing side-wall quenching (SWQ). The 2D SWQ case is well-established [30–32] and includes two essential flame regimes: an unstretched adiabatic flame regime and a non-adiabatic flame quenching at the wall. Even when ignoring unsteady [33] or turbulent effects [34,35] standard flamelet models fail to capture some physics, as discussed by Efimov et al. [36]. Significant modeling efforts were required to develop advanced flamelet models that can accurately predict the pollutant formation, specifically CO [31, 36].

With this background, the objective of this work is threefold:

- to demonstrate the application of a purely data-driven approach for modeling low-order chemistry manifolds in the simulation of the laminar side-wall quenching configuration;
- to identify suitable input parameters for the ML-based manifold without using prior knowledge about the flame physics (applying a sparse principal component analysis [37]);
- to develop a reliable treatment of source terms at the manifold boundaries to ensure boundedness of the thermo-chemical state during the coupled simulation.

The laminar SWQ configuration provides a benchmark for combustion modeling that is sufficiently challenging and therefore suitable for assessing the predictive capabilities of the ML-based approach developed in this study. Analogous simulation results obtained with detailed chemistry and with a flamelet-based model (QFM) serve as reference datasets.

The paper is structured as follows: Section 2 describes the numerical setup of the investigated configurations, namely the 1D head-on quenching (HOQ) and the 2D SWQ. In Section 3, the machine learning methods are outlined. In Section 4, the results for the HOQ and the SWQ cases are analyzed. First the wall heat flux is compared for the HOQ case in order to verify the model. Thereafter, an analysis of the local heat release and the thermo-chemical state for the SWQ configuration is carried out. Finally, conclusions are drawn in Section 5

## 2. Numerical setups

In this section, an overview of the numerical setups is provided. First, the 1D HOQ configuration is addressed, which is used both for the ANN training and the QFM generation. Thereafter, the more complex 2D SWQ configuration is described.

### 2.1. 1D head-on quenching (HOQ)

In the 1D HOQ configuration, a premixed laminar flame propagates perpendicular towards a wall, where it extinguishes due to heat losses. The 2 cm long domain is discretized by an equidistant mesh with 2000 points (resolution of 10 μm) and time integration is realized by a fully implicit backward differentiation formula (BDF). For the initialization, an adiabatic, stoichiometric freely propagating methane-air flame is used with a fresh gas temperature of 300 K. The wall temperature is fixed at 300 K and molecular transport is modeled using the unity Lewis number assumption. Furthermore, the detailed GRI 3.0 mechanism is used [38]. All flamelet simulations are performed with an in-house solver [39], and the setup has been validated previously by Luo et al. [40].

### 2.2. 2D side-wall quenching (SWQ)

Following the setup of previous works [30,32], the simulation domain of the generic SWQ configuration consists of a two-dimensional rectilinear mesh (30 mm x 6 mm) with an uniform cell size of $\Delta = 50$ μm, see Fig. 1. The numerical setup has been validated extensively [30,32] against experimental data [41,42]. The inlet flow is divided into a fresh (5.5 mm) and burned gas section (0.5 mm), where the latter is used for flame stabilization. The fresh gas is initialized with a premixed stoichiometric methane-air mixture at ambient conditions ($T = 300$ K; $p = 1$ atm), while the burned gas consists of hot exhaust gasses at equilibrium conditions. As inlet velocity, a parabolic inflow profile is used and the burned gas velocity is set to 3.81 m/s, compensating for the density difference between fresh and burned gasses. The velocity profile is shown in Fig. 1. At the wall, a constant temperature of 330 K is assumed in accordance with previous studies [32,42]. This represents an additional challenge for the DNN model, which has to capture the preheating at the wall, and this effect must be considered in the training. For the species mass fractions a zero gradient and for the velocity a non-slip boundary condition is applied. At the outlets, zero gradient boundary conditions are applied to temperature, species mass fractions, and velocity. The spatial and temporal discretization scheme, as well as the time step control, were adapted from Steinhausen et al. [32].

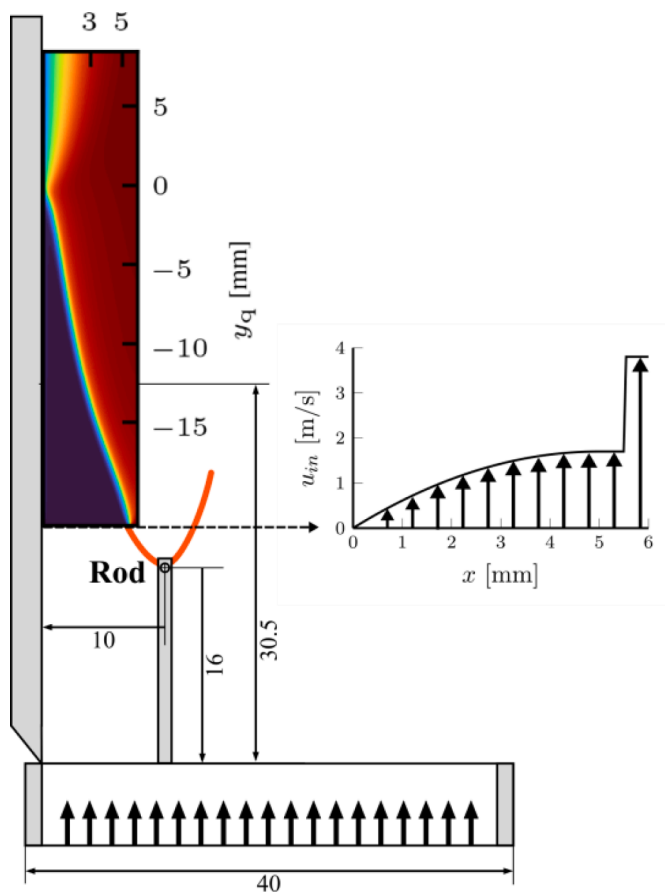

**Fig. 1.** Schematic view of the SWQ burner and the numerical subdomain used for the simulation. All scales are given in mm. The velocity profile at the inlet is also shown.

A detailed chemistry reference solution and corresponding simulation results with a tabulated manifold approach were obtained with in-house solvers based on OpenFOAM (v2006). Different tabulated manifold approaches have been developed and validated for this setup [32], from which the quenching flamelet-generated manifold (QFM) by Steinhausen et al. [32] is selected to serve as a reference model. The QFM is constructed from the same 1D HOQ dataset, which is subsequently used for the training of the neural network, and it is parameterized by the control variables enthalpy and $CO_2$ mass fraction.

## 3. Machine learning methodology

This section gives an overview of the ML methodology employed in this work. First, the generation of the ML training dataset is described. Thereafter, the training parameters, the chosen network architecture, and the method for identifying suitable model inputs are specified. A correction method for the non-linear source terms is proposed and discussed in Section 3.3. Afterwards, the assessment of the ML model accuracy is described and finally, the coupling of the ML model to the CFD code is briefly outlined.

### 3.1. Training data generation

The flamelet-based training data is generated by a transient HOQ simulation, as described in Section 2.1. The left plot in Fig. 2 displays the training data points in the original physical space ($x$) and time ($t$) colored by the CO mass fraction. The isothermal wall is located at the left boundary ($x = 0$ mm). Additionally, this plot shows the writing interval of the flame solution. Initially, the solutions are written at a constant time step, whereas when the flame approaches the wall and heat losses start to occur, the writing interval is then determined by the change of enthalpy at the wall by $1 \cdot 10^4$ J/kg. Additionally, 10 freely propagating flames with varying inlet temperatures ranging from 300 K to 340 K are included in the training dataset. This extension of the manifold is carried out in order to account for elevated enthalpy levels in proximity of the isothermal wall in the SWQ configuration ($T_{\text{wall}} = 330$ K). The same procedure was employed for QFM tables in [32] and is visualized in the right plot in Fig. 2, where the HOQ manifold is displayed in the transformed PV and T space, with the added freely propagating flames located above the red line. It can be observed that the additional flamelets extend the manifold at the upper boundary.

In total, the dataset consists of 310 1D flamelets of different enthalpy levels, resulting in 0.62 million points. Furthermore, the data is randomly split into a training (90%) and a validation (10%) dataset. The latter is used to evaluate the predictive capabilities of the model and to monitor the training process to prevent overfitting. Additionally, all dataset entries are normalized to an interval of [0, 1] (min-max scaling) ensuring similar feature value ranges which promotes convergence of the gradient-based optimization algorithm.

### 3.2. Neural network architecture and training

In order to represent the manifold, which is inherent to the dataset described above, two re- quirements have to be met for the ML approach: (1) a suitable ANN architecture has to be chosen together with a training algorithm, and (2) proper input quantities (i.e., features, or control variables) have to be identified.

Here, the neural network architecture of a Dense Neural Network (DNN) is chosen. Fig. 3 shows the structure of a DNN, which is defined by several sequential layers of neurons. Table 1 summarizes the chosen hyperparameters for the ML training, which is carried out using the PyTorch library [43]. The learning rate is reduced by a factor of 5 if the loss does not decrease for 20 epochs.

For the selection of suitable inputs (or control variables) for the DNN, a sparse principal component analysis (SPCA) [37] is performed. Contrary to a regular PCA, for which the principal components are a dense linear combination, i.e. a combination of all thermo-chemical state variables [44], the SPCA algorithm attempts to minimize the number of variables contributing to the principal components, i.e., a sparse linear combination, which makes the result more interpretable. As a result, the main principal components identified from the SPCA only rely on a few of the variables that define the thermo-chemical state. The variables are normalized as described in Section 3.1 prior to the SPCA.

The first three sparse principal components (SPCs) extracted from the training dataset are shown in Table 2 together with their constituents and associated explained variance. The latter describes the ratio of variance contained in the individual sparse principal component compared to the sum of all sparse principal component variances in the dataset, i.e. a measure of the information contained in the variable. The variances of all SPCs sum up to unity. Interestingly, the first sparse principal component (SPC1) resembles a progress variable (PV) consisting of the species involved in the global methane oxidation reaction. Similar combinations of species have been used for the definition of a primary progress variable in the modeling of methane-air combustion [45] and it is emphasized that SPC1 is obtained here without using prior knowledge about the flame physics. The second sparse principal component (SPC2) mainly consists of the temperature ($T$) and negligible contributions of three species. Both, SPC1 and SPC2, are in accordance with the two governing flame regimes contained in the 1D HOQ training dataset: (1) the propagation of a freely propagating flame (characterized by PV), and (2) flame quenching caused by heat losses to the wall (characterized by T). Therefore, PV and T are chosen as model inputs, see Fig. 3. These principal components also agree well with the inputs of the tabulated manifold approaches, where a progress variable and enthalpy have been chosen [32,36]. For convenience, PV is defined as 2 $SPC1 \simeq Y_{H2O} + Y_{CO2} - Y_{O2} - Y_{CH4}$ in the following.

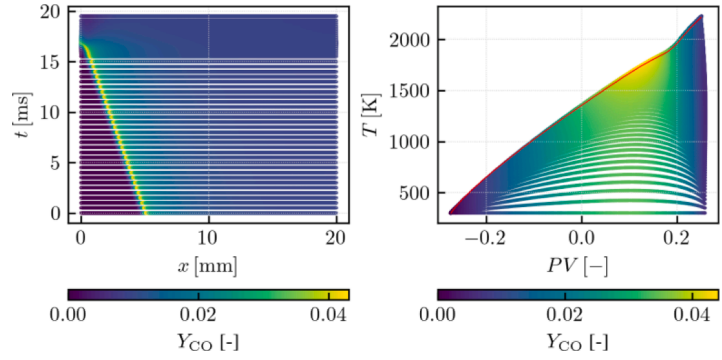

**Fig. 2.** Scatter plot colored by the CO mass fraction for the head-on quenching flame in the physical space ($x$) versus time ($t$) (left). The same dataset is mapped into progress variable ($PV$) and temperature ($T$) coordinates on the right. The additional preheated flamelets, which are incorporated into the manifold, are shown on the right above the red line.

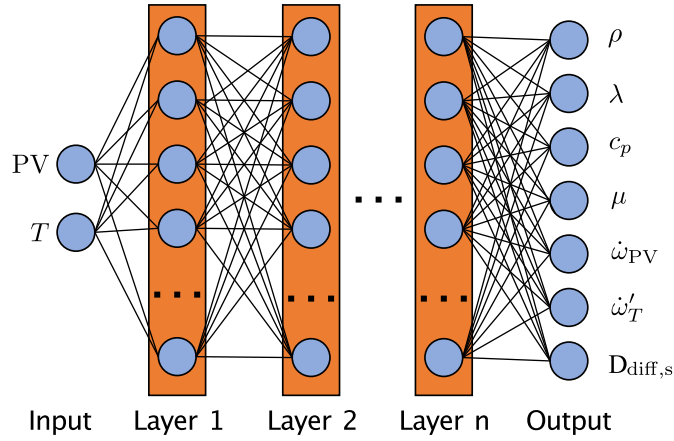

**Fig. 3.** Network architecture of a dense neural network (DNN). The in- and outputs used in this work are shown for the respective layer.

**Table 1**
Hyperparameters for the training of the ML model.

| Hyperparameter | Property |
| --- | --- |
| Epochs | 1000 |
| Batch size | 5000 |
| GPU | 1xTesla K20Xm |
| Optimizer | Adam algorithm |
| Initial learning rate | 0.001 |
| Loss function | Mean-Squared Error (MSE) |
| Activation function | Hyperbolic tangent (tanh) |

**Table 2**
Results of the SPCA for the 1D HOQ training dataset. Only the first three SPCs are shown here. SPC3 is truncated after the first four constituents.

| SPC | Thermo-chemical variables | Variance |
| --- | --- | --- |
| SPC1 | $-0.5Y_{O_2} - 0.5Y_{CH_4}$ $0.5Y_{H_2O} + 0.51Y_{CO_2}$ | 0.365 |
| SPC2 | $-0.004Y_{H_2} + 0.009Y_{O_2}$ $-0.001Y_{H_2O_2} + 1.0T$ | 0.121 |
| SPC 3 | $-0.503Y_{C_2H_6} - 0.492Y_{C_3H_8}$ $-0.446Y_{CH_2O} - 0.335Y_{C_2H_4}$ $+ \ldots$ **(truncated)** | 0.113 |

In the coupled simulation, SPC1 and SPC2 are solved in addition to the velocity and pressure in the employed PIMPLE algorithm, a combination of the SIMPE and PISO algorithm [46]. Assuming unity Lewis number diffusion for all species, the transport equations for the SPC1, i.e. progress variable (PV), and the SPC2, i.e. temperature (T) read

$$\frac{\partial \rho Y_{PV}}{\partial t} + \frac{\partial}{\partial x_j}\left(\rho u_j Y_{PV}\right) = \frac{\partial}{\partial x_j}\left(\frac{\lambda}{c_p}\frac{\partial Y_{PV}}{\partial x_j}\right) + \dot{\omega}_{Y_{PV}}, \quad (1)$$

$$\rho c_p \frac{\partial T}{\partial t} + \rho c_p \frac{\partial}{\partial x_j}(u_j T) = \dot{\omega}'_T + \frac{\partial}{\partial x_j}\left(\lambda \frac{\partial T}{\partial x_j}\right) - \rho \frac{\partial T}{\partial x_j}\underbrace{\left(\sum_{k=1}^N c_{p,k}\left(-D\frac{\partial Y_k}{\partial x_j}\right)\right)}_{D_{\text{diff,s}}}, \quad (2)$$

where $\rho$ is the density, $u$ the velocity, $D$ the diffusion coefficient, $\dot{\omega}_{Y_{PV}}$ the progress variable source term, $c_p$ the heat capacity, $\lambda$ the heat conductivity, and $\dot{\omega}'_T$ is source term of the temperature, i.e. the heat release rate (HRR), respectively. The term $D_{\text{diff,s}}$ in Eq. (2) represents the temperature diffusion caused by the species diffusion. During the numerical solution of Eqs. (1) and (2), PV and T are used as inputs for the employed DNN to retrieve the thermochemical quantities highlighted in Eqs. (1) and (2) (marked by gray boxes) as outputs from the DNN. For the training process, the number of layers and neurons was varied, until a network architecture with three hidden layers containing 64 neurons each has proven sufficient to accurately represent the thermo-chemical variables included in the chemistry manifold. Furthermore, a different training strategy, namely the training of one DNN for each output, was evaluated without observable improvement on the accuracy or computational performance for the manifold investigated here. It is emphasized that this could be different for other cases.

### 3.3. Source term modeling

The accurate description of the chemical source terms can be challenging for ML approaches based on neural networks. Specifically, in a transient simulation small, but non-zero, source terms for fresh gas conditions or equilibrium conditions on the burnt side of a flame (i.e. thermo-chemical states at the boundaries of the manifold) can lead to an error accumulation for the transported control variables (PV, T). Subsequently, the transported variables can drift to unphysical states outside the training interval, where the predictions by the DNN degrade and exacerbate the drift further. Due to their inherent, gradient-based optimization, neural networks asymptotically approach but cannot reach perfect accuracy in regression tasks, which is required here to accurately capture the flame physics. This aspect can easily be missed in the a-priori validation step, since the error at the boundary is small.

Previous works proposed to train additional networks for subsets of the manifold [8,12,25-27]. This can decrease the error made at the manifold boundaries to an acceptable threshold, but it does not completely solve the issue and further introduces computational overhead since the appropriate model has to be identified for every computational element in every time step at simulation runtime.

Here it was found enough to set the source terms to zero close to the PV boundaries. The predicted output of all source terms, namely $\dot{\omega}_{PV}$ and $\dot{\omega}'_T$ is then conditioned on the progress variable at simulation runtime:

$$\dot{\omega} = \begin{cases} 0 & \text{if } \overline{PV} \leq 0 \\ \text{DNN(PV, T)} & \text{if } 0 \leq \overline{PV} \leq 1, \\ 0 & \text{if } \overline{PV} \geq 1, \end{cases} \quad (3)$$

where $\overline{PV}$ is the scaled progress variable. $\overline{PV} = 0$ and $\overline{PV} = 1$ correspond to $PV_{min} = -0.275$ and $PV_{max} = 0.271$ respectively. These values are given by the manifold and are constants in this case, because only one mixture fraction level is considered. Generally, $PV_{min}$ and $PV_{max}$ could also be defined as a function of the mixture fraction if required. The above constraint is realized by a mask vector, which is described in more detail in the next section. It ensures that the numerical solution of the input values of the DNN stays bounded to the known training interval. It is noted that this measure can be easily applied to arbitrary inputs of the DNN.

### 3.4. Model validation

In order to measure and quantify its accuracy, the ML model is tested with the validation dataset (not included in the training). The prediction quality is measured with the coefficient of determination defined as

$$R^2 = 1 - \left[ \sum_{i=1}^{N} (\Phi_i - \widehat{\Phi}_i)^2 \right] \left[ \sum_{i=1}^{N} (\Phi_i - \bar{\Phi})^2 \right]^{-1}$$

where $\Phi_i$, $\bar{\Phi}$, $\widehat{\Phi}$, and $N$ are the true value, the mean value of $\Phi$, the network's prediction, and the number of samples, respectively. For the ML model employed here, all model outputs reach an $R^2$ score of 0.999.

Particularly, CO has been investigated and identified as a relevant quantity for flame-wall interactions in previous studies [30,32,36]. Fig. 4 allows to assess the overall predictive capabilities of the tabulated manifold (QFM) and the ML model for the CO mass fraction in comparison to the two reference datasets (DC) of HOQ and SWQ. Extracting the inputs from the DC datasets, the QFM and ML model are utilized to predict the CO mass fraction which is then compared to the DC references (a-priori analysis). Optimal predictions are located on the diagonal line in Fig. 4, where the predicted CO is identical to the reference DC value. It can be observed that both the QFM and ML model show high accuracy for the HOQ and SWQ datasets. Only slight overestimates are observed for the SWQ for both the ML and QFM model, as indicated by few gray points above the diagonal. This over-prediction of CO is caused by a substantially (factor of 2) higher heat loss to the wall in the HOQ configuration (i.e. the training data) compared to the SWQ configuration. This difference can be attributed to the different directions of the temperature gradients in the respective cases [32,36,47]. Overall, the ML model yields satisfying results and is applied to a coupled simulations of both reference configurations next.

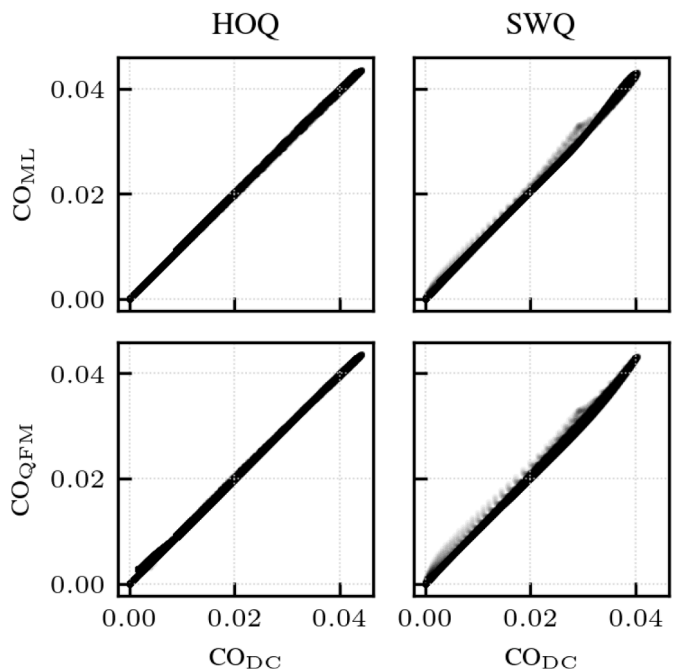

**Fig. 4.** A-priori analysis of tabulated manifold (QFM) and the machine learning model (ML) on the detailed chemistry dataset of head-on quenching (HOQ) and side-wall quenching (SWQ) for the mass fraction of CO.

### 3.5. Model coupling

Before coupling the model to the CFD code two post-processing steps are performed:

1. In order to apply the network in the CFD, the scaling and rescaling operation of in- and outputs are added to the exported network.
2. All source terms are corrected based on the input of the progress variable according to Eq. (3).

The first step allows the generic case-independent implementation of the ML interface into the CFD code. As outlined in the previous section, the source term correction is realized by means of a mask vector, that determines at runtime, which cells contain a thermo-chemical state located at the boundary of the manifold. Subsequently, the output vector of the source term is modified for positive values of the mask vector. Finally, the DNN is coupled to the OpenFOAM-based solver via the PyTorch C++ API for usage in the coupled simulations.

## 4. Results and discussion

In this section, two cases are simulated using the previously described ML model to represent the manifold. First, the model is verified on the 1D HOQ configuration, which was also used to generate the training data. Second, the model is applied to the laminar SWQ configuration described in Section 2.2. By this means, the predictive capabilities of the ML model are assessed for a flame-wall interaction case which is different from the training database.

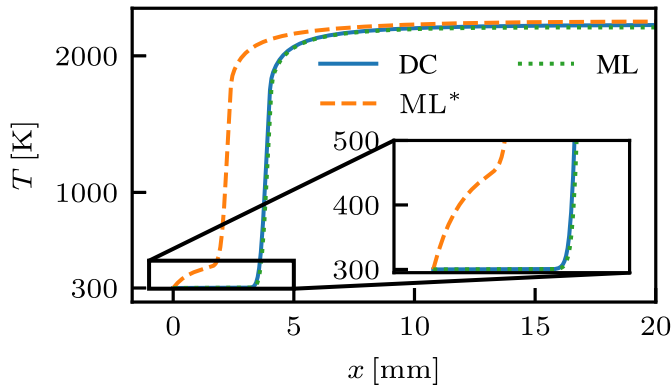

**Fig. 5.** Snapshot of coupled HOQ simulations with a DC and two machine learning models with (ML) and without the source term correction (ML*) according to Eq. (3).

### 4.1. Results for the 1D HOQ configuration

A snapshot of the coupled HOQ simulation with (ML) and without the source term correction (ML*) is shown together with the DC result in Fig. 5. The ML* model predicts small, non-zero temperature (and progress variable) source terms in the preheat zone of the flame, which accumulates to an unphysical temperature increase over the simulation runtime. At the beginning of the simulation, ML* underpredicts the flame propagation speed by approximately 30%, which later turns to an overprediction of similar order of magnitude after a considerable temperature increase. This effect can hardly be identified from the a-priori analysis and it depends on the initialization and duration of the simulation, and the absolute prediction error by the ML model for source terms at the manifold boundaries. In comparison, the ML model, which utilizes the source term correction described in Section 3.3, accurately describes temperature profile and flame speed (< 1% deviation compared to the DC reference).

Furthermore, Fig. 6 shows the wall heat flux for the ML model, the tabulated manifold (QFM), and the DC reference result. It is found that both ML and QFM models recover the overall trend shown by the DC reference result, but overpredict the wall heat flux at the quenching point, characterized as the point in time of maximum wall heat flux, by a similar order of magnitude. It is thereby verified, that the ML model yields comparable results to a tabulated manifold approach, generated from the same dataset used in the training of the neural network.

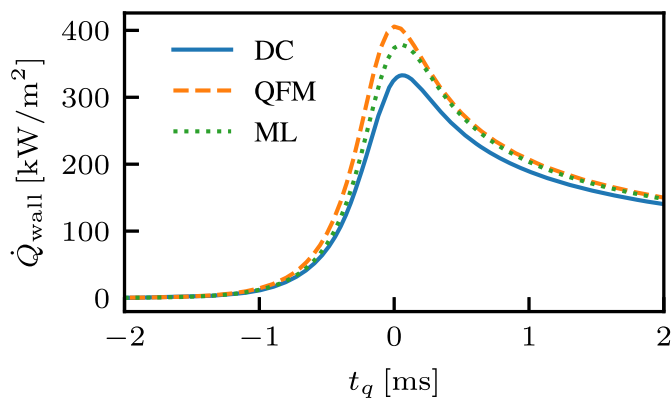

**Fig. 6.** Comparison of the wall heat flux obtained for coupled HOQ simulations with a DC model, a tabulated manifold approach (QFM), and a machine learning model (ML), relative to the quenching time.

### 4.2. Results for the 2D SWQ configuration

The coupled simulation results obtained with the ML model for the 2D SWQ configuration are compared with a detailed chemistry and a tabulated chemistry calculation (QFM, c.f. Section 2.2). The numerical results are analyzed with respect to a relative coordinate system that uses the quenching height as the origin of the wall-parallel direction. The quenching height is defined by the maximum wall heat flux [30,32].

Fig. 7 depicts the temperature contour of the SWQ configuration for the ML model as well as the DC and QFM references. The contours of the reduced models show qualitatively good agreement at the quenching point as well as the shape and position of the adiabatic flame branch. The isolines for three temperature levels (310 K, 320 K, and 330 K) indicate the heating of the fresh mixture (300 K) in close proximity of the wall, caused by the slightly elevated wall temperature (330 K). These preheated states are captured by adding preheated adiabatic flamelets to the training dataset (c.f. Section 3.1). Interestingly, the isoline for 330 K for the QFM shows a small deviation at the point where it meets the wall boundary. This difference can be attributed to the transport of enthalpy for the QFM model instead of temperature for the ML model. The transport of enthalpy requires special treatment of the constant temperature wall boundary condition. In this case, a secondary table is used to retrieve the wall enthalpy for a given progress variable and constant temperature. Whereas, for the ML model the temperature is transported and the isothermal boundary condition is directly imposed.

#### 4.2.1. Analysis of local heat release rate

Following previous investigations on SWQ [30,32,36] the local HRR as a global flame property is used to assess the predictive capabilities of the ML model. The 2D contours of the normalized HRR at the quenching point are displayed in the top row of Fig. 8. The HRR decreases considerably within 0.5 mm distance to the wall showing that the flame quenching process occurs very close to the wall. A detailed assessment of quenching distances for the given SWQ configuration is beyond the scope of this work. Here the reader is referred to Zirwes et al. [47], who investigated the influence of several factors (e.g., chemical mechanism, diffusion model, etc.) on the quenching distance of a SWQ configuration. Overall, a qualitatively good agreement of the HRR contour of the ML model with the DC and QFM references can be observed.

To allow a quantitative comparison, HRR profiles are extracted at three different distances from the quenching point ($y_q = 0$), indicated by the horizontal white dotted lines in Fig. 8. The extracted profiles are shown in Fig. 9. At the three quenching heights $y_q$, both, the peak position and value, of the HRR profiles obtained with the ML model align well with the DC reference solution. Small deviations can be observed at the quenching height $y_q = 0$ mm, where the peak of the ML model is slightly shifted towards the wall and for $y_q = 0.5$ mm, where the peak is slightly overestimated. However, the model yields similar results as the QFM, demonstrating that it is able to accurately capture the local HRR.

#### 4.2.2. Analysis of the thermo-chemical state

As previously mentioned, CO has been identified as a relevant quantity for flame-wall interactions in previous works [30,32,36]. The CO mass fraction and the temperature fields, which are obtained for the SWQ configuration by the DC, QFM and ML modeling approaches, are analysed next.

Contours of the CO mass fraction around the quenching point are displayed in the bottom row of Fig. 8. Here, it can be observed, that the CO production has its maximum within the reaction zone of the flame. The high CO concentration in the quenching zone at the wall indicates an incomplete combustion, where further oxidation of CO was no longer possible.

The local thermo-chemical state is further analysed by comparing the CO mass fraction over temperature at different wall distances in Fig. 10. The locations are indicated by the vertical white dotted lines drawn in Fig. 8. An adiabatic freely propagating premixed flame is additionally included as a reference. Close to the wall at $x = 0.1$ mm significant differences between all the simulations and the adiabatic flame can be observed, indicating an incomplete combustion process. When moving away from the wall ($x = 0.5$ mm and $x = 2$ mm), the CO profile shifts from a quenching state due to heat loss to the wall to an almost adiabatic state.

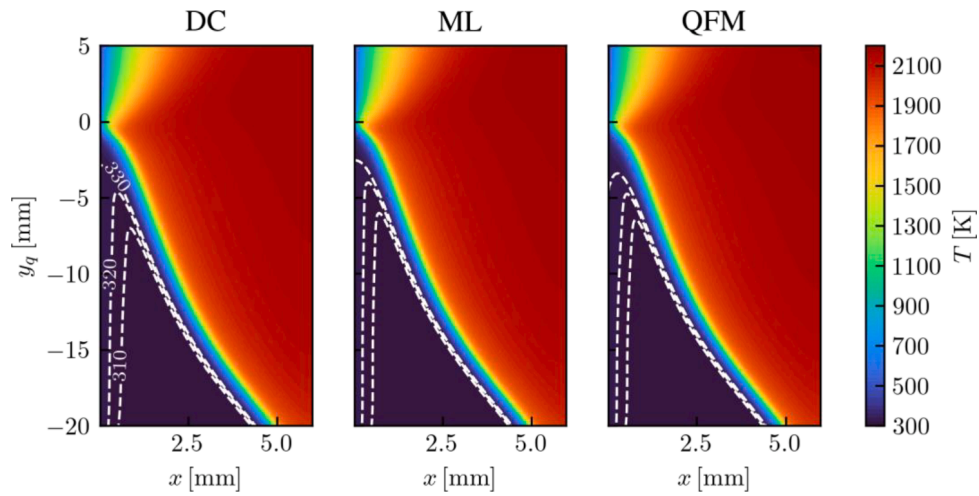

**Fig. 7.** 2D contours of the temperature for the SWQ configuration obtained from simulations with detailed chemistry (DC, left), with the machine learning model (ML, middle) and with the tabulated manifold (QFM, right). Isolines for three temperature levels (310 K, 320 K, and 330 K) are added to highlight the heating of the fresh gas mixture (300 K) by the slightly elevated wall temperature (330 K). Note that the domain is cropped to aid the visual inspection of the quenching point.

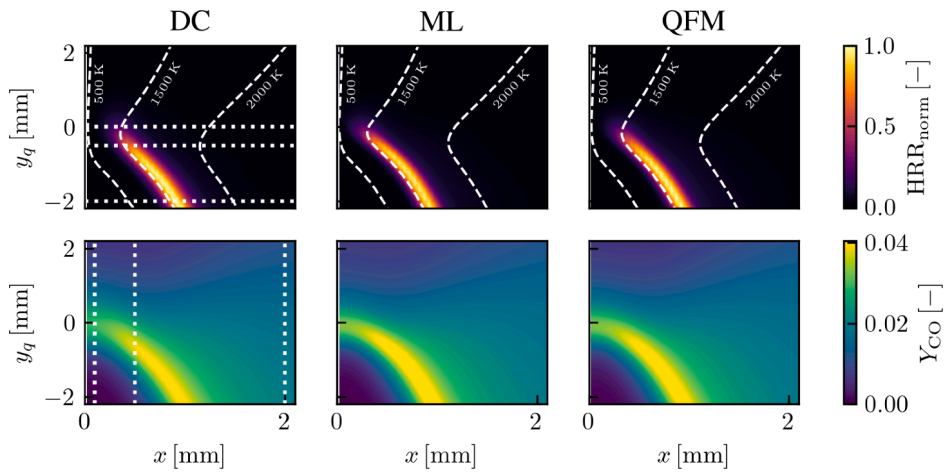

**Fig. 8.** 2D contours of the normalized local HRR (top) and CO mass fraction (bottom) for a detailed chemistry (DC, left), a tabulated manifold (QFM, right), and a machine learning (ML, middle) model in proximity of the quenching point for the SWQ configuration. In the upper row additional isolines of temperature $T$ are shown (dashed white lines).

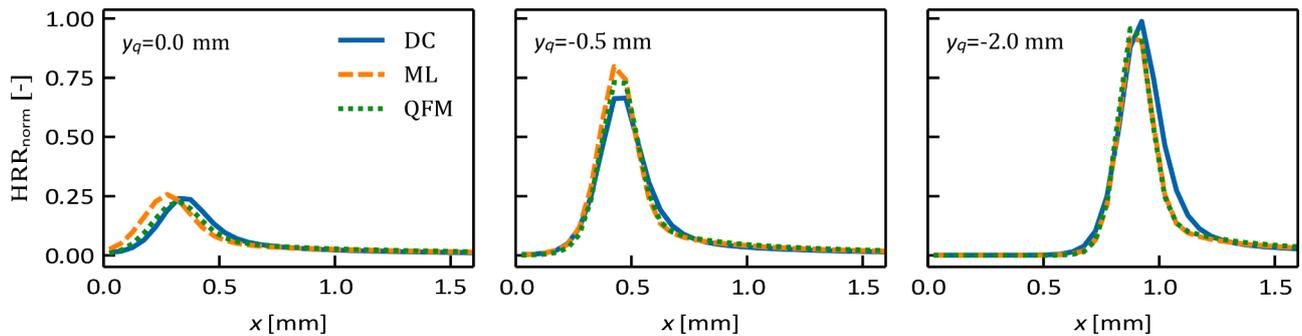

**Fig. 9.** Comparison for the HRR of the SWQ of the detailed (DC), tabulated (QFM), and machine learning (ML) chemistry. The profiles are extracted along wall-normal lines at different heights $y_q$, which are displayed as white solid lines in Fig. 8.

A slight over-prediction of CO can be observed for both the ML and QFM profile. This is in agreement with the conclusions made by the a-priori analysis (c.f. Section 3.4). This over-prediction can be attributed to the different rates of heat transfer to the wall in HOQ and SWQ [32, 36,47]. To improve the prediction of CO, Efimov et al. [36] introduced an additional reactive control variable to account for this effect of varying rates of heat transfer, which could be added to the ML approach in future works. The ML model adequately predicts regions close and far to the quenching point, accurately accounting for the heat losses. Overall, the results show good agreement with the DC reference solution and are qualitatively and quantitatively comparable to the results obtained with the tabulated manifold model.

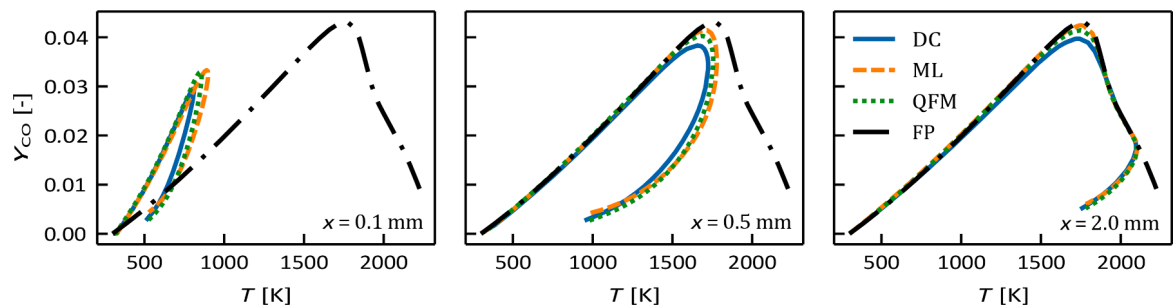

**Fig. 10.** Comparison for the thermo-chemical state of the SWQ of the detailed (DC), tabulated (QFM), and machine learning (ML) chemistry at different axial positions. The profiles were extracted from the vertical white solid lines in Fig. 8. Additionally, the thermo-chemical state of an adiabatic premixed flame (FP) is added as a reference state.

## 5. Conclusions

A data-driven approach has been presented that includes the parameterization and the training of a machine learning model representing a low-order chemistry manifold. The model is coupled to a CFD solver and utilized for the 2D simulation of a premixed methane-air flame undergo- ing side-wall quenching. With an emphasis on the ML modeling techniques, procedures for the selection of suitable input parameters (based on a sparse PCA) and an efficient method for the correction of non-linear source terms at the manifold boundaries have been demonstrated. It was shown in the case of a 1D head-on quenching flame how the accumulation of errors caused by the incorrect prediction of source terms at the manifold boundaries causes an unphysical increase of the transported control variables, which eventually leads to an incorrect flame speed. Contrary, the results of the ML model, which includes the proposed correction method, showed good agreement for flame speed and wall heat flux with the detailed chemistry reference solution. Subsequently, the model was applied to a generic SWQ configuration, where its predictive capabilities for the local heat release rate and the CO production near the wall were analyzed. The ML model showed comparable results to the tabulated manifold approach in comparison to the detailed chemistry reference results. This underlines the ability of ML approaches to capture complex combustion phenomena accurately, such as flame-wall interaction.

In summary, ML chemistry models based on neural networks provide a promising alternative to the conventional approach of manifold tabulation, compensating for some of its shortcomings. The DNN requires only 2% of the QFM's memory, while the computational cost remains similar. However, the performance heavily depends on DNN architecture [48] and hardware, which will be the subject of future studies. This work can serve as the basis to investigate more complex flame configurations in future works, involving aspects such as differential diffusion and stretch effects, turbulent combustion, or hydrogen/hydrocarbon fuel blends.

**Declaration of Competing Interest**

The authors declare that they have no known competing financial interests or personal relationships that could have appeared to influence the work reported in this paper.

**Data availability**

Data will be made available on request.

**Acknowledgments**

The work was supported by the Graduate School Computational Engineering and by the Grad- uate School of Energy and Science at the Technical University of Darmstadt. This work has been partially funded by the Deutsche Forschungsgemeinschaft (DFG, German Research Foundation) – Project Number 237267381 – TRR 150, by the project "Center of Excellence in Combustion", which received funding from the European Union's Horizon 2020 research and innovation pro- gram under grant agreement No° 952181, and by the Federal Ministry of Education and Research (BMBF) and the state of Hesse as part of the NHR Program.